\documentclass[sigconf,authorversion=true,nonacm=true]{acmart}

\usepackage{geometry}
\usepackage{booktabs}
\usepackage{balance}
\usepackage{tikz}
\usetikzlibrary{arrows,arrows.meta,calc,colorbrewer,decorations.pathreplacing,external,fit,matrix,patterns,shadows,shapes,shapes.misc}
\tikzexternaldisable%

\usepackage{todonotes}
\presetkeys{todonotes}{inline}{}

\usepackage{pgfplots}
\usepgfplotslibrary{groupplots,colormaps,colorbrewer,statistics}
\usepackage{pgfplotstable}
\pgfplotsset{compat=1.15}

\usepackage{soul}
\usepackage[absolute,showboxes]{textpos}

\usepackage{pifont}

\usepackage{xspace}
\newcommand{\eg}{\textit{e.g.,}~}
\newcommand{\ie}{\textit{i.e.,}~}

\newcommand{\one}{(\textit{i})\xspace}
\newcommand{\two}{(\textit{ii})\xspace}
\newcommand{\three}{(\textit{iii})\xspace}

\makeatletter
\renewcommand{\paragraph}[1]{\vspace*{0.03in}\noindent{\bf #1.}\hspace{0.25ex \@plus1ex \@minus.2ex}}
\newcommand{\paragraphb}[1]{\vspace*{0.03in}\noindent{\bf #1}\hspace{0.25ex \@plus1ex \@minus.2ex}}
\makeatother

\begin{document}

\title[Reliable Firmware Updates for the Information-Centric Internet of Things]{Reliable Firmware Updates for the\\ Information-Centric Internet of Things}

\author{Cenk G{\"u}ndo\u{g}an}
\affiliation{%
  \institution{HAW Hamburg}
}
\email{cenk.guendogan@haw-hamburg.de}

\author{Christian Ams\"uss}
\affiliation{%
}
\email{christian@amsuess.com}

\author{Thomas C. Schmidt}
\affiliation{%
  \institution{HAW Hamburg}
}
\email{t.schmidt@haw-hamburg.de}

\author{Matthias W{\"a}hlisch}
\affiliation{%
  \institution{Freie Universit\"at Berlin}
}
\email{m.waehlisch@fu-berlin.de}

\renewcommand{\shortauthors}{C. G{\"u}ndo\u{g}an et al.}

\begin{abstract}
Security in the Internet of Things (IoT) requires ways to regularly update firmware  in the field. These demands ever increase with new, agile concepts such as security as code and should be considered a regular operation. Hosting massive firmware roll-outs present a crucial challenge for the constrained wireless environment.
In this paper, we explore how information-centric networking can ease reliable firmware updates. We start from the recent standards developed by  the IETF SUIT working group and contribute a system that allows for a timely discovery of new firmware versions by using cryptographically protected manifest files.
Our design enables a cascading firmware roll-out from a gateway towards leaf nodes in a low-power multi-hop network.
While a chunking mechanism prepares firmware images for typically low-sized maximum transmission units (MTUs), an early Denial-of-Service (DoS) detection prevents the distribution of tampered or malformed chunks.
In experimental evaluations on a real-world IoT testbed, we demonstrate feasible  strategies with adaptive bandwidth consumption and a high resilience to connectivity loss when replicating firmware images into the IoT edge.
\end{abstract}



\maketitle

\setlength{\TPHorizModule}{\paperwidth}
\setlength{\TPVertModule}{\paperheight}
\TPMargin{5pt}
\begin{textblock}{0.8}(0.1,0.02)
     \noindent
     \footnotesize
     If you cite this paper, please use the ICN reference:
     C.~G{\"u}ndo\u{g}an, C.~Ams\"uss, T.~C.~Schmidt, M.~W\"ahlisch. Reliable Firmware Updates for the Information-Centric Internet of Things. In \emph{Proc. of ACM ICN}, ACM, 2021.
\end{textblock}

\section{Introduction}\label{sec:intro}

 The deployment of Information-Centric Networking (ICN)~\cite{adiko-sind-12,xvsft-sinr-14} on embedded wireless devices~\cite{olg-ccnte-10} was first considered a decade ago. With the advent of the Internet of Things (IoT), early experiments~\cite{bmhsw-icnie-14} confirmed benefits for constrained multi-hop networks from operating  NDN~\cite{jstp-nnc-09,zabjc-ndn-14} as a network layer directly on top of data links. Since then a large body of work has proposed and evaluated ICN in the IoT context~\cite{bmhsw-icnie-14,pf-britu-15,acim-icnis-15,sblwy-ndnti-16,arxpp-kip-17,szsmb-avdir-17,gklp-ncmcm-18, caamr-uisfb-18,adbbm-rdndn-20}. Essential findings show that hop-wise forwarding with caching and a leaner network stack improve network performance over IP as well as the adaptability to lossy regimes.

Regular software updates are part of the common life cycle for today's computer systems, and increasing security and agility demands require a similar practice for the IoT. The distribution of firmware or application updates on the Internet, however, is one of the most challenging and resource-consuming tasks~\cite{mfpa-dcrbi-09}. Major updates from popular vendors are repeatedly visible as peak loads at Internet exchange points. Consequently, it is natural to question whether the constrained, lossy networks of the IoT can carry such burdens and how update campaigns may perform.

\begin{figure}
  \centering
  \includegraphics{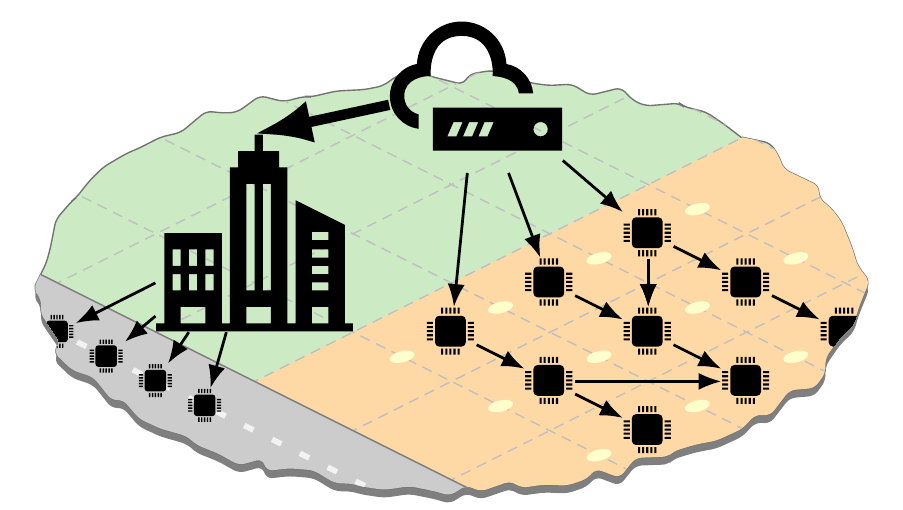}
  \caption{Massive firmware roll-out campaign in distributed and heterogeneous networks}%
  \label{fig:scenario}
\end{figure}

In this work, we devise and evaluate procedures for  reliable, secure, yet scalable software updates in the constrained IoT. Our target objective is the massive roll-out of firmware in edge networks as visualized in Fig.~\ref{fig:scenario}. We want to prove feasibility by leveraging the potential benefits of ICN forwarding and caching.
Our key contributions are \one  context-specific naming, version discovery, and verification, \two  scalable and reliable chunk distribution across updating network nodes with inbuilt DoS detection, \three  thorough experimental evaluations of different update strategies in a real testbed with realistic multi-hop radio links. 

In this paper, we can show that reliable updates of large firmwares in deep multi-hop topologies are indeed feasible. Our findings indicate that firmware dissemination in large networks nested up to seven hops complete within 10 to 30 minutes. We also observe that rapid roll-outs in these networks will fully exhaust resources, whereas slower, cascading update strategies leave sufficient resources at intermediate nodes for continued operations.

The remainder of this paper is structured as follows. We introduce the problem of firmware propagation in the IoT along with related work in Section~\ref{sec:related-work}. Section~\ref{sec:fwupdates} presents the core concepts of secure and reliable firmware updates. A thorough evaluation and discussion of the results follows in Section~\ref{sec:evaluation}. Finally, we conclude with an outlook in Section~\ref{sec:conclusion}.


\section{The Problem of Firmware Propagation and Related Work}\label{sec:related-work}

\subsection{Challenges in low-power regimes}
Secure firmware roll-out campaigns for large-scale IoT deployments demand a coordinated interaction with great regularity between multiple stakeholders.
Vendors prepare and publish firmware versions and local site administrators oversee the roll-out procedure.
An autonomous firmware update without physical proximity can drastically reduce the roll-out time and management overhead for local site administrators.
IoT devices connect through low-power and lossy networks (LLNs) to powerful border routers.
Especially in industrial and rural settings where infrastructure is challenged by natural and regulatory constraints, wireless multi-hop networks are prominent and continuous access to deployed hardware is not always feasible.
These regimes are subject to radio interferences and individual link error probabilities that accumulate in a destructive manner.
In addition, limited maximum transmission units as well as low bandwidth and high delay link capabilities further complicate the distribution of large firmware objects, which necessarily split into hundreds to thousands of fragments.
While corrective actions on the link, network, and application layer usually recover packet loss, small amounts of retransmissions behave additive and induce link stress in broadcast range, which impacts energy expenditures of battery-operated devices.
The exhaustive task of delivering image files also opens up significant attack vectors for denial of service (DoS) attempts.
Willfully tampered or inadvertently modified firmware images deplete network and memory resources to a point where devices neglect mission-critical duties.

The importance of well-thought-out firmware roll-out architectures that efficiently operate in low-power regimes and display a resilient security posture is undisputed.
Several approaches have been proposed in research or have already been deployed in industrial solutions.

\subsection{Firmware updates in the IoT}
SUIT~\cite{RFC-9019} is a recent addition to the menagerie of firmware update architectures.
It is driven by the eponymous IETF working group and aims for a standardized update mechanism in constrained IoT networks that is reliable and secure.
SUIT specifies a concise and machine-processable manifest document~\cite{draft-ietf-suit-information-model-08,draft-ietf-suit-manifest-11} that describes meta-data of firmware images, such as their download location, firmware version, and optional processing steps to decompress and decrypt binaries.
This architecture relies on the Internet protocol stack for retrieving updates and therefore expects certain protocol mechanisms to be present, like congestion and flow control, packet fragmentation, and the ability to resume corrupted transfers.
Given its current momentum at the IETF, we consider SUIT as a suitable blueprint for our information-centric firmware update approach.

ZigBee~\cite{za-zs-15} is a protocol specification harboring various network solutions to inter-connect a wide range of heterogeneous, ZigBee certified devices.
It builds on IEEE~802.15.4 and is prominently used by several product lines, such as Philips Hue, OSRAM lightify, and some Xiaomi devices, albeit not always securely~\cite{rswo-igncz-17}.
In the ZigBee Over the Air (OTA) Upgrade Cluster module, clients regularly poll firmware information, or a server performs \textit{Image Notify} push operations for clients not in hibernation.
The distribution of upgrade images via broadcast or multicast is not recommended due to a missing point-to-point security.
In this case, ZigBee advises a separate unicast attestation with the upgrade server after completing an image transfer.
In contrast to ZigBee, we believe the vendor-independent manifest files of SUIT to concisely organize meta-data provide a greater accessibility to the update process in heterogeneous network deployments.

\begin{figure*}
  \centering
  \includegraphics{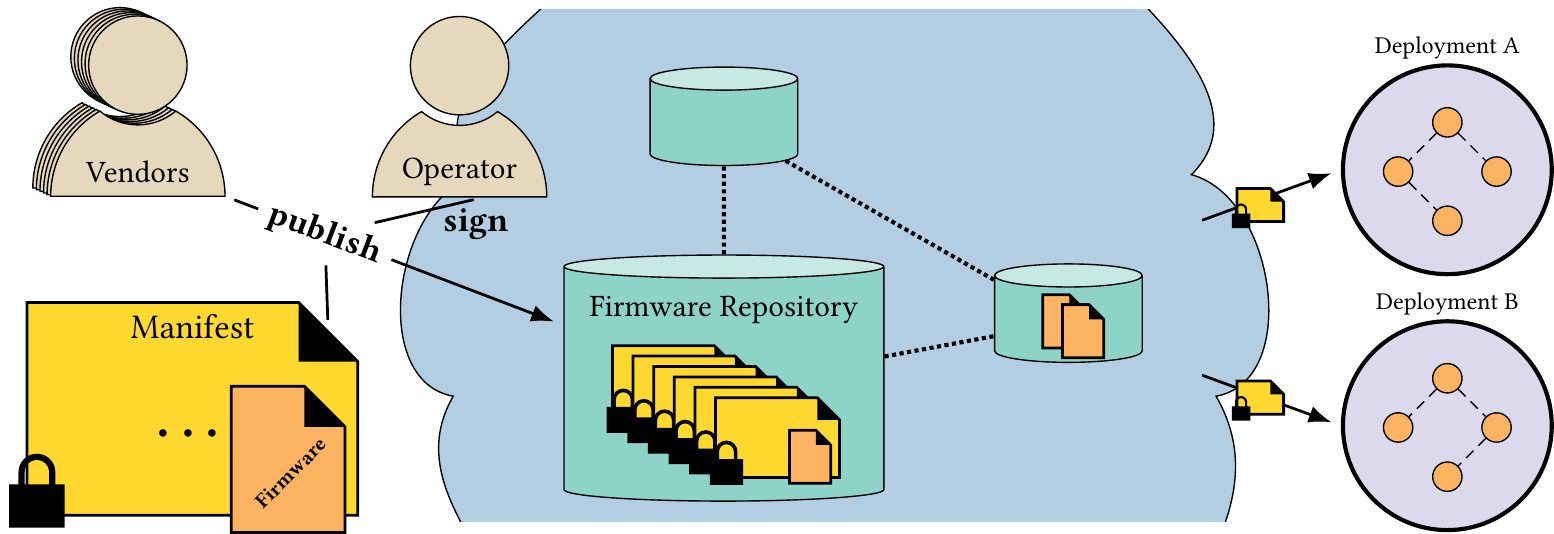}
  \caption{Overview on the back-end system of our information-centric, reliable firmware roll-out approach.}%
  \label{fig:architecture}
\end{figure*}

\subsection{Reliable content transfers and data management in constrained networks}
Large data objects, such as uncompressed binary images with moderate software complexity for embedded devices, can reach file sizes in the range of tens to hundreds of kilobytes.
Prior to the IoT era, wireless sensor networks (WSNs) explored network reprogrammability of low-power devices in broadcast media and reliably disseminated large data objects (\eg a firmware) using epidemic routing methodologies~\cite{khb-npdiw-02,hc-dbddp-04,she-rcumw-03}.
Delicate adjustments to the classic flooding, such as node density awareness, windowing, the use of negative acknowledgments (NACKs), and unicast requests with broadcast data transmissions have shown promising results in lossy networks.
Due to the generally inconsistent protocol layering in former WSNs, packets exceeding link MTUs in constrained network environments had to be fragmented on the application level.

In contrast, the current IoT mostly builds on IPv6 and to bypass transmission limits of typical link layers, the IETF designed 6LoWPAN~\cite{RFC-4944}---a convergence protocol to adapt IPv6 functionalities to challenging LLNs.
It supports a header compression to reduce header verbosity and a fragmentation scheme~\cite{RFC-4944}, which caps at 2048 bytes and is therefore inoperable for firmware propagations.
The constrained application protocol (CoAP)~\cite{RFC-7252} is part of the IETF envisioned IoT network stack and supplements IoT networks with a RESTful communication paradigm.
Block-wise transfer~\cite{RFC-7959} is an add-on to CoAP for splitting payload into equally sized blocks, which are then iteratively transmitted with minimal server-side state.
Chunking on the CoAP level further enables the use of CoAP reliability features for each separate block.

Recent studies~\cite{aarl-raini-19,gklp-ncmcm-18,mwt-tucin-16} reveal a superior data delivery performance for named-data networking (NDN)~\cite{jstp-nnc-09} in low-power networks compared to end-to-end IoT protocols, such as CoAP and the message queuing telemetry transport for sensor networks (MQTT-SN)~\cite{mqttsn12}.
NDN leaves the fragmentation of larger named-data objects to upper layers, since naming decisions for newly created chunks are highly application-specific.
Link fragmentation extensions~\cite{sz-nlpd-12,draft-mosko-icnrg-beginendfragment-02,gksw-dlcli-20} operate below NDN and modify the packet structure.
Other approaches~\cite{gb-nrvnd-15,bgnt-sieoc-13,draft-mosko-icnrg-ccnxchunking-02} apply a fragmentation and naming scheme on the application to yield a structured access to data chunks with predictable names.


\section{Building Blocks for Reliably Updating Firmware with NDN}\label{sec:fwupdates}

\subsection{Roll-out campaign management}

We design a secure and reliable campaign management system for firmware roll-outs that handles the delivery of software updates to numerous constrained edge devices in multiple sites using NDN\@.
We use the SUIT~\cite{RFC-9019} model as a blueprint for our information-centric approach and adopt essential system components and the same terminology.
\figurename~\ref{fig:architecture} illustrates our name-based back-end proposal, which consists of three components: \one publishing and versioning firmware images and manifest files by vendors, \two managing the storage of chunked software updates by an operator and providing access to the IoT deployment sites, and \three a timely notification of version updates and a reliable delivery of necessary updates towards edge devices on the IoT side.

\subsection{Firmware preparation and publication}

\begin{figure}[H]
  \centering
  \includegraphics{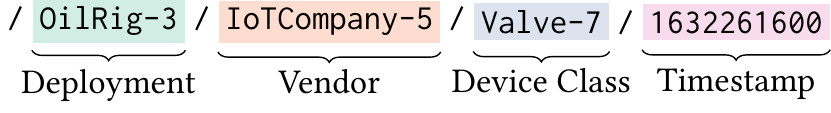}
  \caption{Namespace schema.}\label{fig:namespace}
\end{figure}

\paragraph{Namespace management}
Large site deployments can consist of heterogeneous devices from varying vendors and the highest level of interoperability is essential to construct an energy-efficient system.
A systematic namespace management regulates all interactions between vendors and IoT devices.
\figurename~\ref{fig:namespace} demonstrates our name schema used for all components, ranging from upper-layer application functions down to forwarding and caching duties.

Each deployment has a globally unique name and may identify an offshore drilling rig, segments of a connected urban network, or a smart home environment.
We consider the deployment identifier as the leading component in our name schema to keep forwarding states towards single deployment sites minimal, \ie they most certainly aggregate due to the spatial proximity of devices within the scope of a deployment.
Vendor names are equally globally unique like deployment identifiers and both components are managed by the same, external registry.
Finally, a device class designates a specific firmware for all nodes of the same product type.
The timestamp component describes the actuality of a firmware and is encoded as a Unix timestamp with a predefined granularity.
To fully leverage the in-network caching abilities of NDN, binaries are prepared for device classes instead of yielding unique binaries for each single device.
This also reduces the binary management overhead on the vendor site.

\paragraph{Firmware generation}
Vendors precompile firmware images for their deployed product lines and keep track of the software versioning.
Since binaries are prepared for device classes, the images cannot ship with sensitive data.
Device-specific configurations are rather obtained on run-time after a successful firmware installation on an IoT node.
This requires that each IoT device is provisioned with vendor-specific data for bootstrapping purposes during the manufacturing stage or with the use of an out-of-band channel, which is already common practice for real-word deployments.
This data outlasts firmware upgrades and is stored independently of the program code, \eg in a dedicated address space on the flash memory, or using an SD card.
To protect the firmware integrity, vendors also generate a message digest of the binary alongside the firmware image.

\paragraph{Preparation of firmware chunks}
The small-sized MTUs in common network link technologies disallows the transmission of images in single network packets.
For a successful delivery, an image fragmentation at the vendor and a reassembly at the IoT edge devices is necessary.
Fragmentation on convergence layers~\cite{RFC-4944,gksw-dlcli-20} is a solution to provide a hop-wise, fragmented delivery between two peers, but due to the layering, these schemes make the caching of individual fragments impossible.
Thus, we focus on a fragmentation approach that chunks the image on application level and reassembles them at the IoT edge device after all chunks have been successfully retrieved.

The reassembly of fragmented images must be as simplistic as possible for the constrained devices.
We therefore follow a linear chunking of the image file in our solution, where each chunk is of fixed length (the last chunk being an exception).
The reconstruction on the low-power devices is straightforward as fixed-length chunks can be joined using offsets, which makes the need for an ordered delivery unnecessary.
Chunk sizes may vary between device classes, since different link-layers will yield different MTUs.
Each chunk is addressed by appending a monotonically increasing chunk identifier \textit{/chunk/id} to the base name (see \figurename~\ref{fig:namespace}), starting at \textit{/chunk/0}.

\begin{figure}
  \centering
  \includegraphics{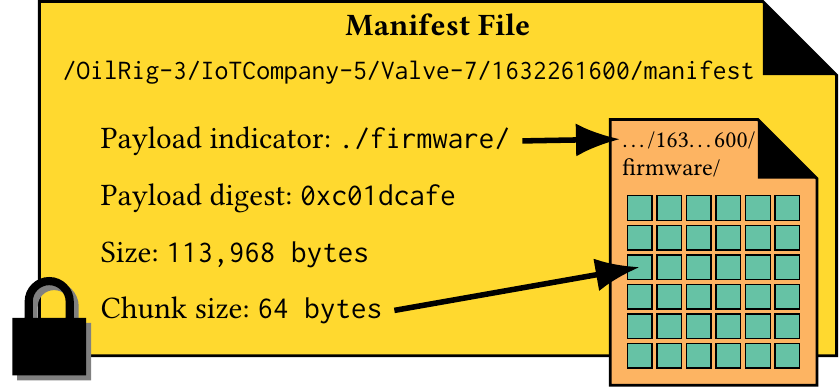}
  \caption{Manifest description and fixed-length chunks of a corresponding firmware image.}%
  \label{fig:manifest}
\end{figure}

\paragraph{Manifest description}
As demonstrated in \figurename~\ref{fig:manifest}, a vendor also creates a manifest file following the SUIT model to organize meta-data on the firmware version and binary image.
They include the binary size and message digest as well as parameters for the chunking algorithm.
To preserve the authenticity of manifest files, vendors sign them during the upload process.
This also protects the message digest, which is later used to validate the final firmware image on the IoT devices.
The manifest is addressed using the base name (see \figurename~\ref{fig:namespace}) and the suffix \textit{/manifest}.

\paragraph{Firmware upload and binary management}
Once all artifacts have been produced, a vendor delivers the manifest and firmware chunks to the corresponding deployment operator to serve them in a firmware repository.
The publication process runs in an automated manner and requires an authentication framework to ensure consistency and security, such as the publicly auditable bookkeeping service NDN DeLorean~\cite{yaszz-ndasd-17}.
A firmware repository stores versioned images of all vendors and retains them until they are purged.
For replication purposes, an operator can deploy multiple firmware repository instances, which then synchronize using any data set synchronization solution~\cite{za-lcdds-13,zlw-pespn-16}.

\begin{figure}
  \centering
  \includegraphics{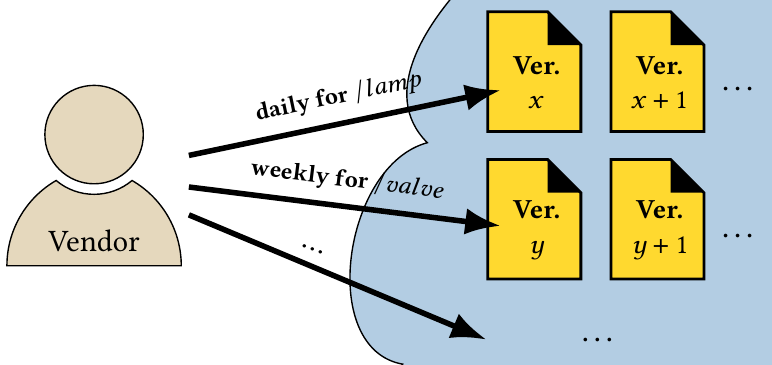}
  \caption{Vendor publishes firmwares and aligns date granularity with polling interval of device classes.}%
  \label{fig:version-date}
\end{figure}

The uploaded firmware binaries and manifests are tagged following the naming scheme in \figurename~\ref{fig:namespace}.
The suffix for the actual image is \textit{/firmware} and the corresponding manifest is \textit{/manifest}; chunks are accessed via \textit{/chunk/id}.
The timestamp in the naming scheme updates for new firmware versions to reflect the upload time and the granularity of the epoch time is coordinated with the polling interval of the devices, \eg a daily alignment on midnight would yield \texttt{1632261600} for \texttt{09/22/2021 00:00:00}.
A vendor chooses different degrees of granularity on a device class level as illustrated in \figurename~\ref{fig:version-date}.

\paragraph{Discussion: full versus incremental updates}
We design our firmware roll-out approach to always deliver the full binary.
An alternate approach would explore the use of differential algorithms to compute software differences, \eg with \textit{bsdiff}~\cite{p-ndec-03}, and transmit them in the form of binary patches.
While it is undemanding for powerful firmware repositories to calculate a minimal diff representation, the patch size can grow very quickly for compiled binaries.
Especially in the IoT, binaries are compiled with optimizations to reduce the binary size as far as possible to fit the image on the programmable flash memory.
This can lead to large differences for small changes between software versions due to code re-organizations, up to the point, where caching them in the content store becomes unfeasible and would evict application data.

Incremental updates using the linear chunking approach is another alternative, which appears to be attractive on first sight.
Chunks from previous versions could be reused with a correct name mapping in the manifest file to reorder the fragments independent of their sequential chunk identifier.
However, this can quickly inflate the size of the meta-data itself and may require a separate fragmentation for the manifest file.

Sophisticated linking techniques that use auxiliary information about the structure of deployed software modules~\cite{kp-riler-05} can produce concise patches compared to na\"ive diff algorithms which operate on the byte level.
Run-time relinking of software components directly on the sensor nodes can lead to minimal diff representations, but requires an extensive tooling support during binary compilation and software installation~\cite{mglms-ffecu-06}.
While we only focus on the propagation of the binary, the actual representation of such an artifact (full binary or an increment) is rather secondary and may only necessitate slight protocol adaptations regarding the naming schema.

\subsection{Firmware update process}

\paragraph{Firmware version discovery}
IoT devices are naturally provisioned with an up-to-date firmware version before they become operational in a deployment.
Over time, vendors release new software versions to update device functionalities or to handle security related issues.
Depending on the network availability, a device may be a single or multiple versions behind the current firmware.
A version discovery is therefore the first step to any upgrade process.

Two fundamental strategies exist when determining the availability of a new firmware update: \one proactively notifying the IoT devices using push mechanics, and \two periodically polling the firmware repository.
While timely notifications from a firmware server to the IoT device minimize the operational run time of outdated software components, it also bears the following issues.
First, notifications are not guaranteed to arrive in low-power regimes where nodes favor extended sleep cycles. Second, it requires server-side state and maintenance overhead to keep track of deployed versions as well as topological information to ensure node reachability, and last, the push mechanism is not native to NDN\@.

Our approach primarily relies on a pull-driven version discovery, where the embedded devices periodically request the latest manifest file for a dedicated time frame.
Vendors convey a sensible polling interval on device class level, \eg a daily check on midnight for remotely deployed gas valves, or flexible intervals based on harvested energy levels for battery-less sensors.
These guidelines are programmed during run-time configurations and may change at an operator's discretion.
Since the Unix epoch denotes the actuality of a firmware image in the name schema, all IoT devices need to re-adjust drifting system clocks using an external mechanism, \eg by relying on the time information of an equipped GPS module, or by operating a time synchronization protocol, such as NDNTP~\cite{mm-nndnt-20}.

\begin{figure}
  \centering
  \includegraphics{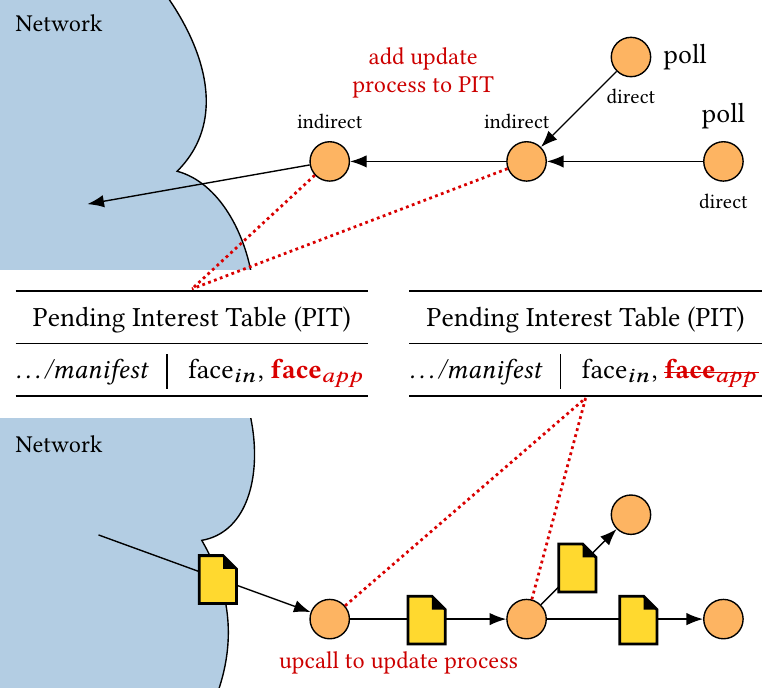}
  \caption{Direct and implicit version discovery.}%
  \label{fig:version-discovery}
\end{figure}

\paragraph{Retrieval of firmware versions}
To discover a new version, IoT devices send Interests to the name that identifies the latest firmware version by setting the correct time frame.
Following our previous example, the Interest may describe the name \textit{/OilRig-3/IoTCompany-5/Valve-7/1632261600/manifest}, in which the requested time frame is greater than the time frame of the locally running firmware.
The firmware repository returns a manifest file if the requested update is available, \ie a vendor published the binary image for the specified time frame.
Interest retransmissions retry the update request for a configurable, but limited number of times to recover manifests from packet loss.
In the event that a requested manifest does not exist yet or all corrective actions fail, the Interest times out as part of the default NDN forwarding logic and the IoT node triggers a subsequent update request on the next polling interval, potentially on midnight of the next day.
Negative acknowledgments for Interests (NACKs) is a supported NDN protocol element to hint at the absence of requested data or to carry nuanced error codes of the application.
For our retrieval mechanism, they may include application-level indications about the latest firmware version.
While NACKs are not necessary to ensure a continuous operation, this feature \one provides an optimization to reduce the amount of retransmissions when polling for a new, non-existent firmware version, and \two assists with the convergence of updates for devices that missed a version publication, \eg due to network unavailability.
The lifetimes of cacheable NACKs need to be aligned with the firmware release cycles to prevent them from wrongly satisfying requests for eventually released versions.
To reduce the attack surface, NACKs require similar security considerations as manifest packets.

IoT nodes may be disconnected for longer periods from the core network and thereby may fall several versions behind.
Fixing a maximal update frequency at the application level allows a node to always request the latest version at the appropriate Unix epoch. Hence, outdated devices need not attempt to retrieve obsolete versions. 
Forwarding states are handled by an external routing system, \eg~\cite{haazz-nnlsr-13,swbw-lcnhp-16}, preferably using a single default route from all IoT devices toward the firmware repository.

\paragraph{Implicit consumption of firmware versions}
Polling intervals of devices within the same class can drift apart over time, so we utilize an implicit version discovery process to reduce the amount of individual manifest requests and to increase the reactivity of the firmware roll-out. 
Each IoT forwarder in a multi-hop request path compares incoming manifest requests with its own device class.
On a positive match and if the requested name has a greater epoch time than the currently operating firmware, then the update process of a forwarding device internally registers to the same entry in the Pending Interest Table (PIT) as illustrated in \figurename~\ref{fig:version-discovery}.
This assures that each device of the same class on a request path consumes the manifest and then initiates the retrieval procedure of the firmware image before its local request interval triggers.

\begin{figure}
  \centering
  \includegraphics{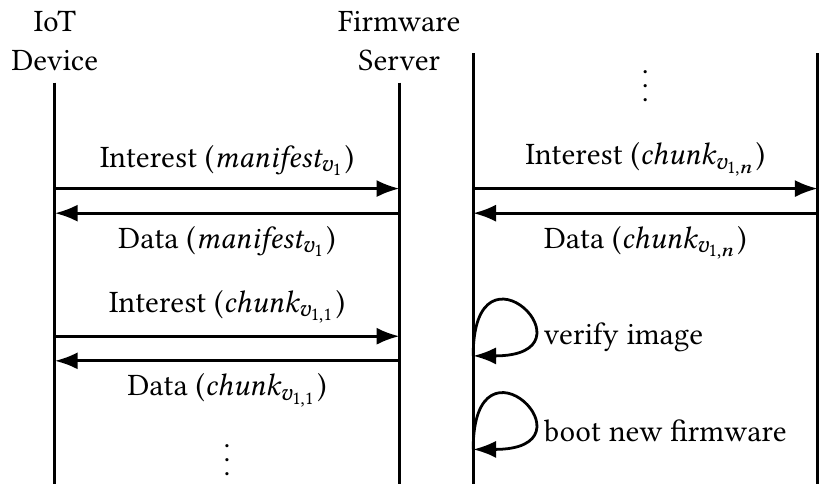}
  \caption{Iterative retrieval of firmware chunks.}%
  \label{fig:seq-iot}
\end{figure}

\paragraph{Retrieval of firmware image chunks}
Once an edge node determines the need for a version upgrade by receiving an up-to-date manifest file, it prepares for retrieving the associated binary image.
Initially, the manifest signature is validated using key materials previously provisioned by a vendor.
On a failed check, the upgrade process aborts and this incident is reported to the vendor.
A valid signature triggers the retrieval of all firmware chunks as designated by the manifest.
Each chunk is addressed by appending the chunk identifier (\textit{/chunk/id}) to the base name, where \textit{id} starts at 0 and gradually increments to the maximum chunk number as appointed by the manifest.
This also ensures that a few resources are available for alternative forwarding duties.
Since memory and network resources are generally limited in low-power regimes, the system uses a stop-and-wait automatic repeat-request (ARQ) error-control method, \ie each chunk is retrieved iteratively as illustrated in \figurename~\ref{fig:seq-iot}.
Characteristically, resource-constrained class~2 devices~\cite{RFC-7228} equip less than 100~KiB of main memory, where larger parts are inevitably consumed by the operating system, the network stack, and reserved for application purposes.
This leaves only persistent memory components, \eg flash and SD cards, to buffer intermediate chunks during the retrieval.
In contrast to the available RAM, external memory often displays storage capacities that are orders of magnitude larger, but uncoordinated access can also consume the available energy budget as I/O operations tend to be slower and energy-draining.

We define two different chunk retrieval strategies that we assess in our experimental evaluations.
The first method allows concurrent firmware updates from nodes on the same request path.
The second retrieval method disallows overlapping updates and rather prefers an ordered update that cascades downstream into the IoT network.

\begin{figure}
  \centering
  \includegraphics{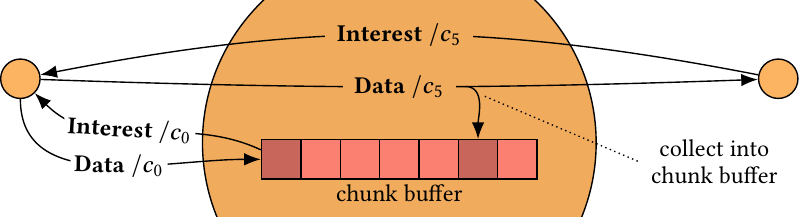}
  \caption{Local buffer collects chunks from overlapping upgrade processes.}%
  \label{fig:buffer}
\end{figure}

\paragraph{Concurrent firmware updates}
While nodes request one chunk at a time, they still perform forwarding duties for other devices.
Overlapping upgrade processes may also yield incoming data objects that are farther advanced in the firmware buffer than the local chunk identifier as illustrated in \figurename~\ref{fig:buffer}.
In this case, a firmware consumer diverts matching chunks with higher progression into the local buffer.
Simultaneously, this buffer is also used for serving incoming chunk requests from other devices.
Although this optimization results in an unordered data retrieval, the use of fixed-length chunks eliminates the need for reorganizing the fragments when reconstructing the image.
Power demanding I/O operations to persistent memory are thus minimized.

\paragraph{Cascading firmware updates}
In this retrieval method, a node denies the delivery of firmware chunks for the same device class as long as a node did not complete the update process itself.
Downstream nodes run into request timeouts for the first chunk and application retransmissions retry the retrieval using a configurable polling interval.
With this strategy, firmware versions propagate hop-wise from a gateway device towards any leaf node of a multi-hop network.

\paragraph{Firmware verification}
After completing the retrieval, all necessary chunks reside in the local chunk buffer and this also concludes the full image reassembly.
A node calculates a message digest across the buffer and validates it against the previously received firmware digest.
On a positive verification, the binary is copied to the correct flash region, the temporary chunk buffer is cleared, and the bootloader is notified to invoke the new firmware.
The timer for the next version request is armed as soon as the new image boots successfully.
Following the SUIT~\cite{RFC-9019} philosophy, the update process still keeps the old firmware image on the device as a backup in case the recent firmware update breaks the node operation.
At worst, the bootloader initiates a fail-safe to re-flash the old binary and return to a correct and consistent behavior.

\paragraph{Firmware replication on connectivity loss}
Once the upgrade completes, a device can also serve the latest manifest and binary chunks to downstream devices.
The advantage of using a linear binary chunking is that an up-to-date forwarder device serves chunk requests directly from its read-only flash region where the currently running firmware resides, without separately consuming main memory.
A firmware version can therefore cascade downstream into the IoT network in a hop-by-hop fashion without necessary operations from the firmware server.
This design confines chunk retrievals to a single link and therefore leads to a reduction in bandwidth usage.
It also provides a loose coupling, so that upgrade processes become resilient to uplink outages and are unaffected by temporary network disruptions.

\begin{figure}
  \centering
  \includegraphics{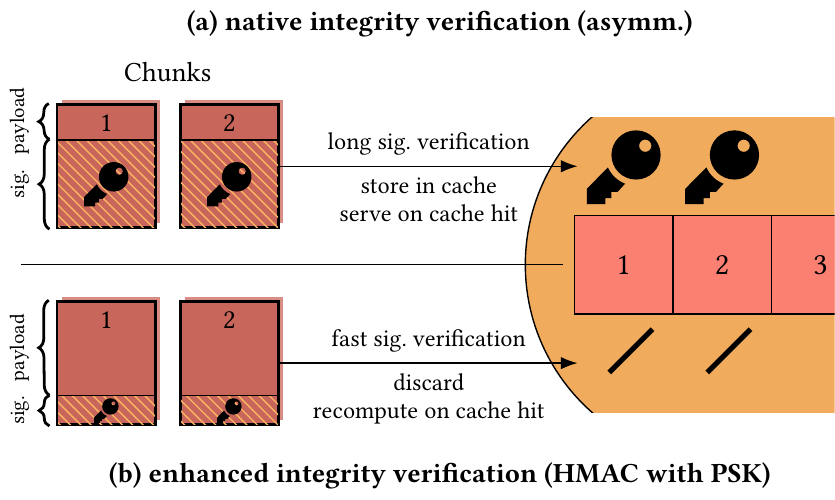}
  \caption{Enhanced chunk-wise integrity verification to save device and network resources compared a native NDN protection with asymmetric cryptography.}%
  \label{fig:chunk-verification}
\end{figure}

\paragraph{Early denial of service (DoS) detection}
Images may consist of hundreds or thousands of chunks, depending on the firmware complexity and the (usually small) MTUs of underlying link-layer technologies.
NDN protects singular content objects (see \figurename~\ref{fig:chunk-verification}a), but \one the chunk-wise computation of digital signatures using asymmetric cryptography is infeasible for the constrained environment, in particular if no hardware acceleration is available~\cite{kblsw-pscli-21};
\two full-length signatures inflate each packet, thereby immensely reducing the actual goodput of the firmware delivery, and \three IoT devices must store message signatures alongside the respective data to serve requests from the local cache.
This consumes a storage capacity that can grow as large as the firmware itself in low MTU scenarios (\eg for 802.15.4 with less than 128~bytes payload room).

\begin{figure}
  \centering
  \includegraphics{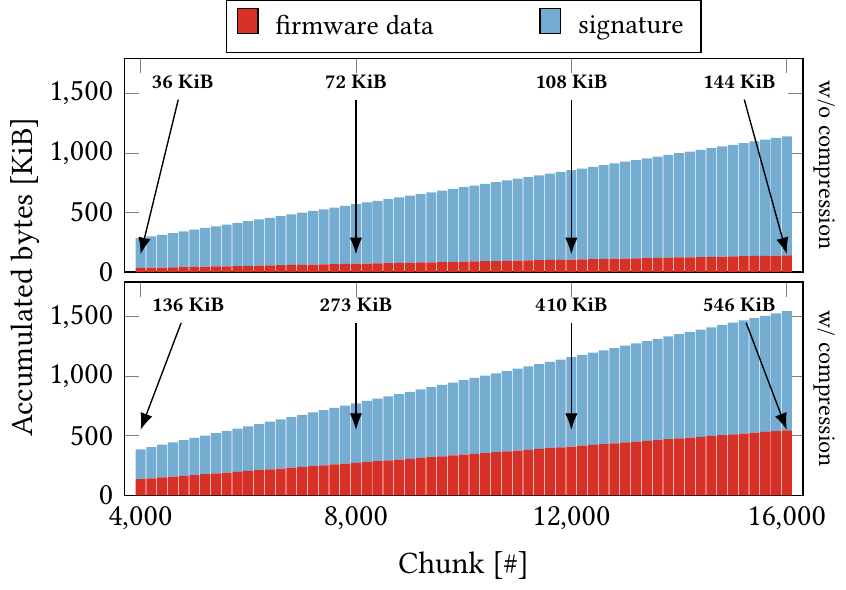}
  \caption{Chunk-wise signature overhead compared to the actual firmware data. Chunks contain 9~bytes (w/o ICNLoWPAN compression) and 35~bytes (w/ ICNLoWPAN compression) of application data.  Signatures are 64~bytes for EdDSA (Curve25519).}%
  \label{fig:chunk-verification-overhead}
\end{figure}

\figurename~\ref{fig:chunk-verification-overhead} illustrates the aggravating effect of comparatively large signature sizes.
In this example, we assume the 802.15.4 MTU, a data name of 16~bytes, a structural NDN encoding overhead of another 16~bytes, and the link-layer header further consumes 23~bytes when using the long MAC address mode.
This sums up to 55~bytes and leaves 73~bytes for the payload and signature.
The Edwards-Curve Digital Signature Algorithm (EdDSA)~\cite{RFC-8032} is a prominent choice in the IoT as it provides a high performance and relatively small signatures of 64~bytes---at least with the Ed25519 curve.
Yet this reduces the available space for application data down to 9~bytes, resulting in numerous chunk packets containing individual signatures.
Even for small firmware sizes of 36~KiB, 4000 chunk transmissions accumulate to a signature overhead of 256~KiB.
For larger images, this linearly increases: a firmware with 144~KiB requires 16000 chunk transmissions and produce a signature overhead of 1~MiB.
The \mbox{ICNLoWPAN} convergence layer~\cite{gksw-dlcli-20} can remove names from Data messages and reduces the structural header overhead.
Following our exercise, these enhancements increase the available space for firmware data from 9 to 35~bytes, thereby requiring four times less chunks to complete the firmware delivery.
Regardless, the signature overhead remains intolerable.
The severity shows in NDN cache environments where each signature has to be stored alongside the chunk data due to the asymmetric aspect of this signature algorithm that prevents IoT devices from generating them.

The integrity and authenticity of a firmware image  is validated against the protected message digest from the corresponding manifest file once all chunks have been received and reassembled. Hence, signatures of individual NDN messages are redundant and we omit for the sake of efficiency. 
Unauthenticated packets, though, open a forceful attack vector to exhaust the resources of the IoT network:
Injecting (even few) illegitimate chunks violates the integrity of the firmware and an identification of these invalid chunks is difficult after firmware reassembly.
The only approach to recover the binary is then to repeatedly request the firmware, which requires all chunks to traverse the network first.
To save device and network resources, a detection of erroneous deliveries and an early exit of the retrieval process is desired.

We augment individual chunks with a keyed-hash message authentication code (HMAC~\cite{RFC-2104}) that is verified upon reception (see \figurename~\ref{fig:chunk-verification}b).
Next to the asymmetric cryptography, NDN already provides the protocol elements to encode a 32-byte HMAC authentication code.
To check for data integrity as well as authenticity, the HMAC requires seeding.
For this early DoS detection module, we assume a pre-shared secret at all devices of a class, which can be pre-installed by the vendor or obtained in an out-of-band manner and eventually protected in secure memory.
A chunk is then recorded in the chunk buffer only after correctly verified by its recipient.
If a chunk validation fails, a recipient repeats requests for invalid chunks only.
After iterated (\eg three) failing verification attempts, a node marks the firmware as irrecoverable, aborts the update process, and notifies the vendor. 

It is noteworthy that these signature hashes based on pre-shared secrets can be discarded during caching in the chunk buffer, since nodes of the same device class can easily re-generate them using the same secret at any time.
This relieves storage capacities, while preserving an intact cache operation for incoming chunk requests.
For low-power regimes with small-sized MTUs, a full HMAC signature may occupy too many bytes in a frame.
For optimization, we only transmit a configurable prefix of the hash, \eg 8, or 16 bytes.
This trade-off increases the susceptibility to hash collisions, but drastically increases the goodput.
Security and robustness of the final image verification remain unaffected by these optimizations.

\section{Experimental Evaluation}\label{sec:evaluation}
In this section, we quantitatively assess our previously outlined information-centric firmware update approach using a real protocol implementation and constrained nodes in a testbed.

\subsection{Experiment setup}
\begin{figure}
  \centering
  \includegraphics{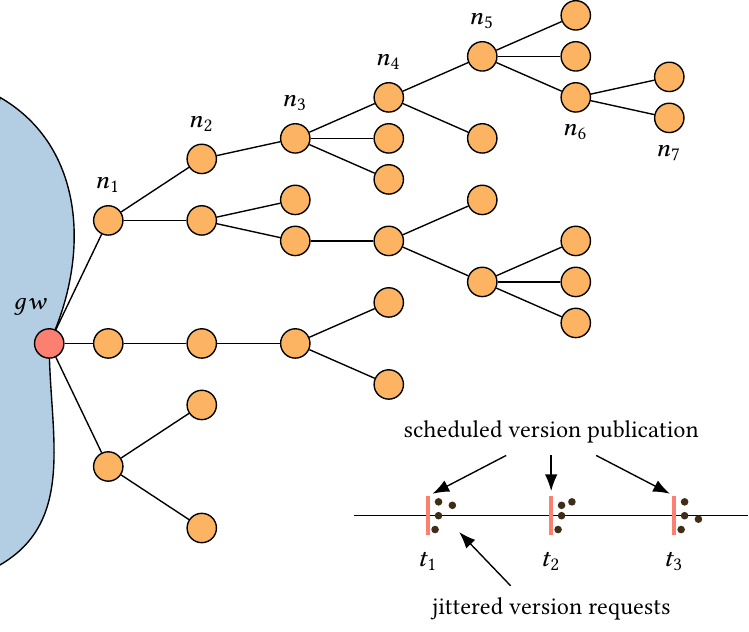}
  \caption{Logical testbed topology modeling multiple branches from rank zero to seven of the forwarding hierarchy.}%
  \label{fig:topology}
\end{figure}

\paragraph{Scenario and network topology}
We conduct our experiments in a wirelessly connected IoT deployment where a gateway node is situated at the network edge to provide an uplink connectivity to a set of 30 IoT devices.
A new binary version is rolled out into the stub network.
On system initialization, the constrained nodes statically arrange in a destination-oriented, directed and acyclic graph (DODAG) as depicted in \figurename~\ref{fig:topology}.
DODAG topologies provide shortest paths from IoT devices to root nodes (\ie gateway or cloud) and therefore incur a minimal routing overhead for the prevalent \textit{converge cast} scenario, \ie a large amount of traffic is directed to or from a central point.
In fact, RPL~\cite{RFC-6550}---the predominant routing protocol for the IoT---uses DODAGs as a fundamental part of its routing system.
While we rely on a static topology in our test environment to sidestep the delays of routing convergence and to solely focus on the propagation of large data objects, an authentic deployment would use an orthogonal routing protocol to dynamically construct and repair the DODAG as necessary.

\begin{figure*}
  \centering
  \includegraphics{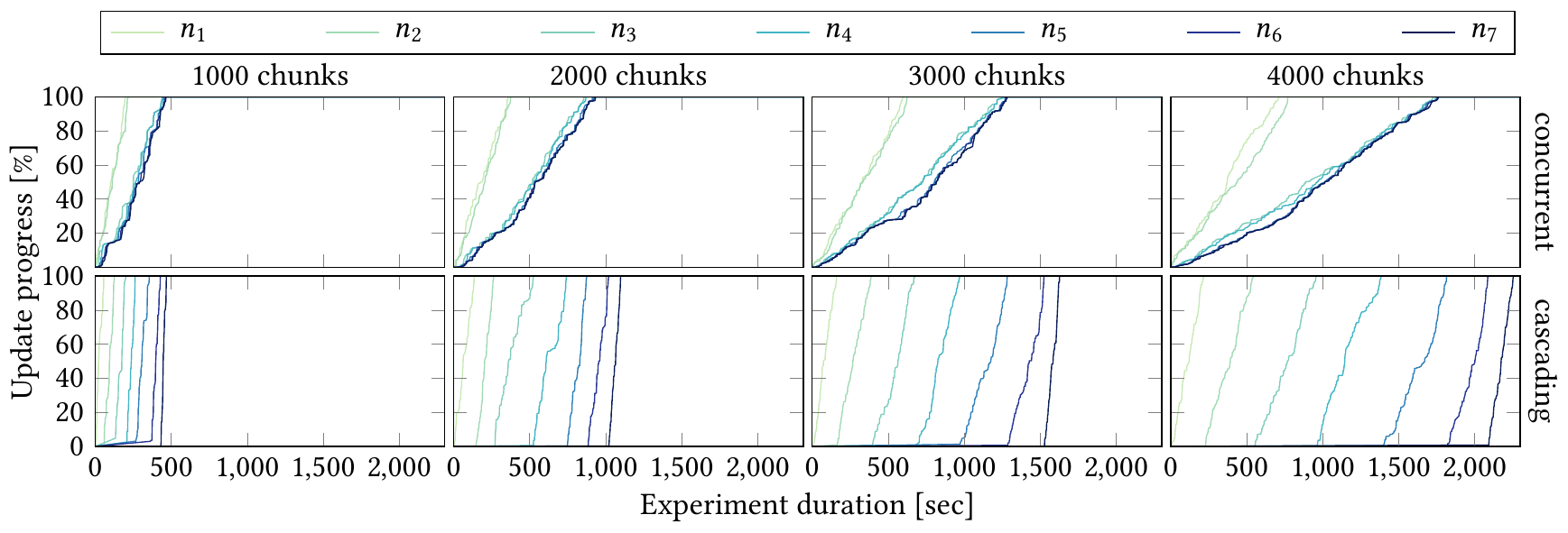}
  \caption{Overall firmware update progression for the selected nodes $n_{1 \ldots 7}$ with an increasing number of maximum chunks using the \textit{concurrent} and \textit{cascading} retrieval strategies.}%
  \label{fig:chunk-retrieval-cumsum}
\end{figure*}

\paragraph{Software and hardware platform}
On all IoT nodes, we deploy the RIOT~\cite{bghkl-rosos-18} operating system in version 2021.04.
It integrates with CCN-lite, which implements a minimal NDN forwarder.
The necessary update logic runs as a small IoT application using the portability layers of RIOT and CCN-lite, which opens the implementation to a wide range of hardware platforms.

We conduct all evaluations on the FIT IoT-LAB~\cite{abfhm-filso-15} testbed.
It features large deployments of several ARM Cortex-M3 based class 2 devices~\cite{RFC-7228} with 64~kB of RAM and 256~kB of ROM\@.
The testbed nodes are equipped with an Atmel AT86RF231~\cite{a-lptzi-09} transceiver to operate on the IEEE~802.15.4 2.4~GHz radio.

\paragraph{Deployment parameters}
We externally align the system clock of the IoT devices and the gateway node with the Unix epoch using the instrumentation tools of the testbed.
In a configured interval of one hour, we generate new binary versions and record the corresponding manifest and image files in the content store of the gateway.
Once the time is synchronized, the IoT nodes request new manifest files from the gateway node as soon as they are generated.
In our experiment, we deploy the same device class throughout the network, \ie the same firmware image for all devices, but also provide a glance at the end of the evaluation on the performance for the other extreme: all nodes are of a different device class.
We separately explore the two retrieval strategies: \textit{concurrent}, where update processes overlap between multiple nodes, and \textit{cascading}, where downstream nodes first wait for upstream nodes to complete the update.

We choose names for manifests and chunks (see \figurename~\ref{fig:namespace}) with a total size of 45~bytes when encoded in the NDN TLV format.
While we increase the image size from 32 to 128~kB in our experimental evaluations, we gradually raise the number of maximum chunks from 1000 to 4000.
Thereby, we keep the chunk size fixed to 32~bytes across all configurations.
This yields a length of 92~bytes for chunk data packets and the total frame size sums up to 115~bytes including the IEEE~802.15.4 link header.
Thus, these parameters produce chunk packets that are very close to the link MTU of 128~bytes.
The NDN forwarder performs three retransmissions in a two-second interval and the application triggers retransmissions in a jittered interval of 10$\pm$5 seconds after a designated chunk request times out.
We configure three link-layer retransmissions that operate in the lower millisecond range, whereby each retransmission is slightly delayed by a random exponential backoff algorithm.

\subsection{Firmware update progress}
In our first evaluation, we gauge the update progression over time for a set of selected nodes with increasing firmware size.
This nodal time measurement starts when the first firmware chunk is requested and terminates on the successful delivery of the last chunk.
Our node selection consists of $n_{1 \ldots 7}$, \ie the nodes that reside on the longest path in our topology.
\figurename~\ref{fig:chunk-retrieval-cumsum} summarizes the various evolutions over the experiment duration.

We observe that both retrieval strategies yield very different progression charts.
In the \textit{concurrent} mode, all update procedures of $n_{1 \ldots 7}$ start almost simultaneously and run concurrently for a designated time.
The first two nodes $n_{1,2}$ advance with a similar chunk retrieval speed in all configurations and the remaining nodes $n_{3 \ldots 7}$ display a similar alignment, albeit with a much slower evolution.
While the firmware distribution with 1000 chunks continues for $\approx$8~minutes to complete for the whole network branch, the duration multiplies to $\approx$30~minutes for an image file of 128~kBytes (4000 chunks).
The \textit{cascading} deployments display the anticipated stop-and-wait characteristic.
Single nodes wait for the immediate uplink node to finish the update process, before any chunk retrievals are invoked.
This serialization positively affects individual update speeds.
In the extreme configuration, the update duration for $n_7$ declines from 30~minutes down to 3~minutes, which is the quickest update completion on the request path.
However, while individual updates appear to be faster in the \textit{cascading} mode, the global roll-out on this path is $\approx$8~minutes slower than in the \textit{concurrent} mode.

\begin{figure*}
  \centering
  \includegraphics{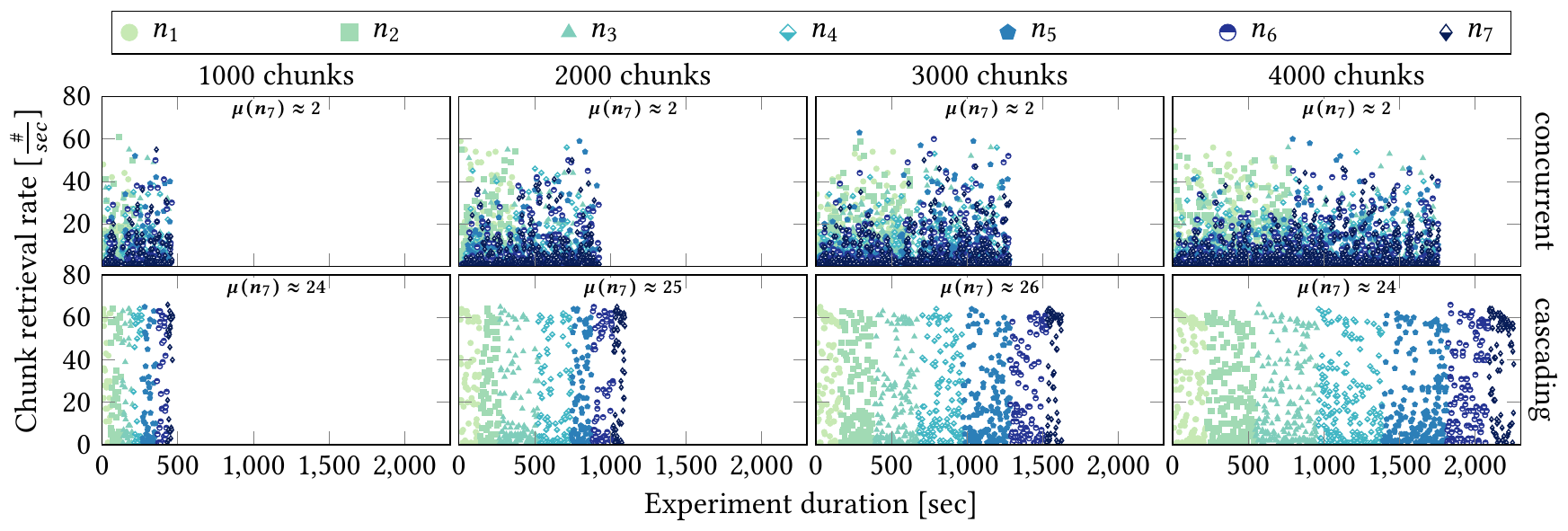}
  \caption{Chunk retrieval rate per second for our node selection using both retrieval strategies.}%
  \label{fig:events-time}
\end{figure*}

\subsection{Goodput analysis}
In our next comparison, we emphasize on nodal chunk retrieval rates to elucidate the previous progression differences.
\figurename~\ref{fig:events-time} displays the amount of accumulated chunks that nodes retrieve in a second.
We observe highly fluctuating rates throughout the update process ranging from zero chunks per second up to accumulated retrievals around 60 chunks per second.
With the \textit{concurrent} retrieval strategy, nodes $n_{3 \ldots 7}$ generally display lower rates while $n_{1,2}$ have ongoing transmissions.
The average performance of the $n_7$ leaf node nets to an average of approximately 2 chunks per second for all configurations.
Roughly at the middle of the experiment duration the first two nodes complete their update process, which leads to slightly increased retrieval rates for the remaining nodes.
This is an indication that nodes in this deployment are competing for bandwidth in the shared wireless medium.
When retrievals are \textit{cascading}, then the number of simultaneously competing nodes in the topology is drastically reduced to single nodes in all request paths of the topology that have overlapping broadcast ranges.
The nodal goodput moderately improves for all nodes across all presented configurations.
For $n_7$, this translates into a performance gain that is nearly twelve-fold.
The evident oscillations are a result of request retransmissions.
Unlike layer 2 retransmissions which operate on the millisecond range and are mostly invisible in the considered timescale, corrective actions on upper layers block the retrieval process by multiple seconds until messages are recovered by the network layer or application, thereby impairing the nodal goodput rates.

\subsection{Link stress}
\begin{figure}
  \centering
  \includegraphics{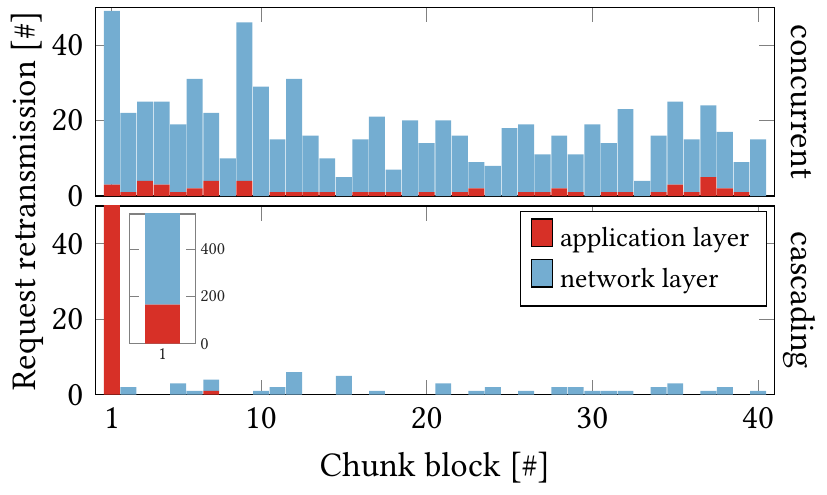}
  \caption{Chunk request retransmissions on the application and network layer grouped into blocks of 100 chunks for the $n_7$ node.}%
  \label{fig:retrans-per-chunk}
\end{figure}

The preceding evaluations suggest that both retrieval methods experience varying degrees of network stress when firmware updates are progressing in the multi-hop topology.
We now measure the link stress for $n_7$ by quantifying the amount of retransmitted chunk requests.
\figurename~\ref{fig:retrans-per-chunk} accumulates request retransmissions for blocks of 100 chunks and differentiates between corrective actions on the network and application layer.
In the \textit{concurrent} configuration, $n_7$ triggers a seemingly continuous stream of $\approx$5--45 retransmissions which is higher at the beginning and then slightly decreases over the experiment duration.
This is in accordance with our former observation that chunk rates increase as soon as competing upstream nodes complete their updates and access to the shared medium lessens.
Overall, the amount of application retransmits is rather minuscule compared to the number of network retransmissions, \ie NDN is able to recover most of the chunks with its three request attempts.

The \textit{cascading} setup shows a much less pronounced retransmission behavior: many chunks  experience no packet loss at all while other groups register less than ten network retransmissions for 100 chunks---still considerably less than the \textit{concurrent} configuration.
This relaxed progression also confirms the previously observed performance gains when firmware images are distributed in a hop-wise fashion.
Application retransmissions are virtually absent, excluding the very first chunk.
$n_7$ retries the retrieval of the first chunk, but $n_6$ denies the delivery until it completes its own update.
This leads to the large amount of $\approx$160 application and $\approx$500 network retransmissions that originate from $n_7$.
These numbers appear to be disproportionately high, however, these packets trigger moderately in the seconds range over a period of $\approx$30 minutes and do not pose a significant stress to the shared medium. Overall, we observe sufficient idle resources to continue regular network operations during the update.

\subsection{Multiparty assessment}
Up until now, our experiments  updates the same device class throughout the network.
A roll-out of a collective firmware image clearly benefits from the NDN multicast support: in-network caches and request aggregations can greatly balance the network utilization.
In this last assessment, we configure a different device class for each device in the deployment to deliver individual binaries to the respective nodes.
While this contrary extreme is usually impracticable in real-world deployments, it gives a sensible estimation on the performance of protocol ensembles without caching and aggregation capabilities.
Due to the low memory, nodes are only able to cache a maximum of 64 foreign chunks, but they mostly evict before they can be utilized by retransmissions, because of the significant chunk flow rate that leads to rapid and frequent cache replacements.
The internal chunk buffer is reserved for the respective binary image of the node and is therefore inaccessible by the remaining nodes.

\begin{figure}
  \centering
  \includegraphics{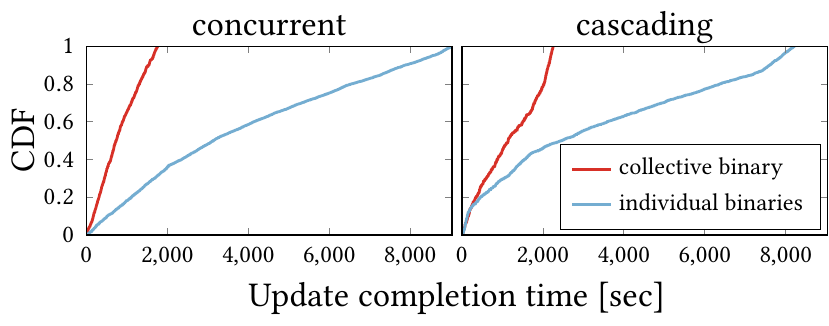}
  \caption{Update completion time for the selected path $n_{1 \ldots 7}$ with a maximum amount of 4000 chunks per firmware image.}%
  \label{fig:chunks-cdf}
\end{figure}

We measure the chunk arrival times for nodes $n_{1 \ldots 7}$ and demonstrate the update progression in \figurename~\ref{fig:chunks-cdf}.
Nodes request 4000 chunks to complete the image delivery, \ie 28k distinct chunks in total are transmitted on that particular path.
The distributions indicate a completion time of $\approx$30 minutes for the setup with a single device class and a collective binary.
On the other hand, the update time considerably decelerates if the NDN multicast features are inactive.
Hence, the update process continues for more than two hours when individual binaries are deployed to propagate.
The missing hop-wise caching ability means that retransmissions need to traverse the full request path up to the gateway node on each retry, which again promotes higher packet loss probabilities due to the generated side traffic for other, ongoing transmissions.
In contrast, in-network caches reduce the number of necessary hops and confine retransmissions in the best case to a single link.
The greater slopes towards the end of both \textit{cascading} measurements are an indication that leaf nodes operate quicker with the coordinated retrieval method due to absent nodes in the vicinity that compete for the bandwidth, irrespective of caching abilities.

\section{Conclusions and Outlook}\label{sec:conclusion}
We have studied massive roll-outs of firmware in large-scale constrained multi-hop  networks, which is an emerging need but also a major challenge  for the IoT edge. We found that  information-centric content replication fosters efficient and reliable chunk dissemination, which makes routinely firmware updates feasible even for nodes that are highly constrained in processing power, memory, and radio capacity. Hop-wise forwarding and in-network caching in particular facilitate update campaigns across homogeneous wireless regimes even with intermittent connectivity.

Using the IETF SUIT  update model as a blueprint, we further devised and evaluated firmware propagation strategies based on the Named Data Networking (NDN) protocol.
We conducted a feasibility analysis using real protocol implementations on a wireless testbed to quantify the effective network performance of retrieving large firmware images in the information-centric Internet of Things.
Our findings indicate that \one a simultaneous, uncoordinated distribution of firmwares results in high cross traffic within the broadcast domain that degrades nodal operability,
\two deployments with collective binaries significantly benefit from in-network caching, and \three a hop-wise, cascading delivery relaxes strain on network resources, allows for continued regular operations during the roll-out process, and preserves limited energy budgets by allowing longer sleep cycles due to prompt firmware deliveries.

This work raises research questions in three directions. First, further insights and optimizations of current design decisions and operational practices are expected to be learned from long-term deployment studies. Second, experiences from massive firmware roll-outs in ICN deployment scenarios may generate valuable feedback for the RESTful, CoAP-centered IoT~\cite{gasw-triwt-20}: Which insights can help to develop the emerging 
data-centric Web of Things? Third, we propose to explore how content object security~\cite{gasw-cosit-21} can be optimized for the IoT to ease voluminous data transfers without sacrificing integrity, authenticity, and DoS resistance.


\appendix
\section*{Acknowledgment}
We want to thank the anonymous reviewers and our shepherd Alex Afanasyev for constructive feedback and inspiration on how to improve the paper.
This work was supported in part by the German Federal Ministry for Education and Research (BMBF) within the projects
\textit{RAPstore -- RIOT App Store} and the Hamburg \textit{ahoi.digital} initiative with \textit{SANE}.

\paragraph{A Note on Reproducibility}
We fully support reproducible research~\cite{acmrep,swgsc-terrc-17} and perform all our experiments using open source software and an open access testbed.
Code and documentation will be available on Github at \url{https://github.com/inetrg/ACM-ICN-2021-FWUPDATE}.


\bibliographystyle{ACM-Reference-Format}
\bibliography{own,rfcs,ids,manet,ngi,iot,internet,layer2,meta,security,theory}

\end{document}